# Fault slip in hydraulic stimulation of geothermal reservoirs: Governing mechanisms and process-structure interaction

Inga Berre[1], Ivar Stefansson[1], and Eirik Keilegavlen[1]

https://doi.org/10.1190/tle39120893.1

## Abstract

Hydraulic stimulation of geothermal reservoirs in low-permeability basement and crystalline igneous rock can enhance permeability by reactivation and shear dilation of existing fractures. The process is characterized by interaction between fluid flow, deformation, and the fractured structure of the formation. The flow is highly affected by the fracture network, which in turn is deformed because of hydromechanical stress changes caused by the fluid injection. This process-structure interaction is decisive for the outcome of hydraulic stimulation, and, in analysis of governing mechanisms, physics-based modeling has potential to complement field and experimental data. Here, we show how recently developed simulation technology is a valuable tool to understand governing mechanisms of hydromechanical coupled processes and the reactivation and deformation of faults. The methodology fully couples flow in faults and matrix with poroelastic matrix deformation and a contact mechanics model for the faults, including dilation because of slip. Key elements are high aspect ratios of faults and strong nonlinearities in highly coupled governing equations. Example simulations using our open-source software illustrate direct and indirect hydraulic fault reactivation and corresponding permeability enhancement. We investigate the effect of the fault and matrix permeability and the Biot coefficient. A higher matrix permeability leads to more leakage from a permeable fault and thus suppresses reactivation and slip of the fault compared to the case with a lower matrix permeability. If a fault is a barrier to flow, increase of pressure because of the fluid injection results in stabilization of the fault; the situation is opposite if the fault is highly permeable compared to the matrix. For the given setup, lowering the Biot coefficient results in more slip than the base case. While conceptually simple, the examples illustrate the strong hydromechanical couplings and the prospects of physics-based numerical models in investigating the dynamics.

## Introduction

Low-pressure stimulation to cause natural existing fractures to slip and dilate, typically termed shear stimulation or hydroshearing, was identified by Pine and Batchelor (1984) as a key mechanism to enhance the permeability of conduction-dominated geothermal systems in basement and crystalline igneous rocks (also called petrothermal systems) (Moeck, 2014). Depending on subsurface conditions, shear stimulation has proven successful for this purpose at several sites (Breede et al., 2013). For example, at the Soultz experimental geothermal site in Alsace, France, the injectivity index (flow rate per unit wellhead pressure under steady-state conditions) for three of the four stimulated wells improved by factors in the range of approximately 15–20. The last well, which had the highest index prior to stimulation, only improved by a factor of about 1.5 (Genter et al., 2010), showing possibly how the preexisting permeability of the formation affects its response to hydraulic stimulation.

Unstable slip along faults manifests as seismicity, which is generally observed related to hydraulic stimulation of geothermal reservoirs in basement and crystalline igneous rocks (Zang et al., 2014) as well as in other subsurface operations involving fluid injection (Ellsworth, 2013). Related to stimulation by hydroshearing, most induced seismicity at geothermal fields has been below magnitude 3, with some notable exceptions where fluid injection has resulted in stress-perturbations that trigger more significant earthquakes. In Basel, Switzerland, stimulation of a granitic geothermal reservoir caused a moment magnitude ($M_w$) 3.4 earthquake. This event occurred five hours after shut-in of the injection well, which was done as a response to preceding increased levels of seismicity. Additionally, three earthquakes of $M_w$ greater than 3 were recorded one to two months after the well was bled off (Deichmann and Giardini, 2009; Bethmann et al., 2011). In Pohang, South Korea, stimulation of the low-permeability crystalline basement reservoir activated a previously unknown fault and induced a $M_w$ 5.5 earthquake almost two months after the final stimulation of the reservoir (Geological Society of Korea, 2019). Analysis of the Pohang event disproved the hypothesis that the largest induced earthquake magnitudes are bounded by the injected volume. Rather, the maximum magnitude in Pohang was sensitive to preexisting tectonic conditions and structures as well as the number of earthquakes induced (Lee et al., 2019). Furthermore, faults can be reactivated far from the injection point (Goebel et al., 2017). For example, the 1967–1968 Denver earthquakes, which were induced by wastewater injection into crystalline rocks beneath the Denver Basin, were mainly triggered 3–8 km from the injection well. The three largest events, up to $M_w$ 4.5–4.8, were triggered one to two years after injection was stopped (Herrmann et al., 1981; Hsieh and Bredehoeft, 1981).

These examples show that fault reactivation and slip represent coupled interactions between induced hydromechanical dynamics and the initial state and faulted structure of the reservoir. While analysis of field data is decisive in understanding the coupled phenomena, the mechanisms resulting in fault slip and reactivations are challenging to study and analyze based on observations. Decameter-scale experiments in a single fault have informed the study of injection-induced slip and seismicity, showing that fluid pressure changes cause aseismic fracture slip and dilation, with corresponding static stress redistribution that in turn can trigger seismic slip (Guglielmi et al., 2015). Hence, while seismicity is associated with slip along faults, slip and permeability enhancement of faults also occur without seismic manifestation. The location of seismic events also has an associated uncertainty (Eisner et al., 2009), which further complicates interpretation of data.

[1]University of Bergen, Department of Mathematics, Bergen, Norway. E-mail: inga.berre@uib.no; ivar.stefansson@uib.no; eirik.keilegavlen@uib.no.

In recent years, physics-based modeling of the dynamics of hydraulic reservoir stimulation has developed substantially. While available data, parameter uncertainty, and model error still limit predictive capabilities, numerical models are important as supplementary tools to understand the coupled process-structure interaction between governing processes — in particular hydromechanical — and the faulted structure of the formation. This requires physically consistent mathematical models and corresponding simulation tools that fully acknowledge the coupled dynamics and dominating structural effects and solve the resulting equations with stable and convergent numerical schemes.

In the following, we discuss reactivation and permeability enhancement of faults as a consequence of low-pressure stimulation. Using simulation studies, we examine and illustrate different mechanisms and characteristics that affect dynamics. We base our model on a formulation of physical principles and constitutive relations, coupling flow in and deformation of explicitly represented faults with flow in and deformation of their surrounding matrix domain, which also incorporates finer-scale structures. For simplicity, the model does not incorporate geochemical and thermal effects, which also may be influential. Geochemistry can, for example, alter fault friction (Wintsch et al., 1995), and cooling leads to thermal contraction and corresponding reduction of fault strength (Ghassemi, 2012; Stefansson et al., 2020).

## Fault reactivation, slip, and dilation

Conceptually, one can distinguish between fault reactivation caused directly or indirectly by pressure changes due to fluid injection (Figure 1). In direct hydraulic reactivation of a fault, increased fluid pressure in the fault — which reduces the effective normal stress on the fault — induces slip (Hubbert and Rubey, 1959). Once a part of the fault is hydraulically reactivated and slips, the immediate stress redistribution can cause further slip along the same fault, indirectly induced by the fluid injection. Indirect fault reactivation can also occur if poroelastic stress changes cause a fault to slip, even if the fault is not at all or only weakly hydraulically connected to the injection region (Figure 1). This occurs, for example, if direct hydraulic reactivation of a fault causes stress changes that reactivate other faults. In hard rocks, slip along a fault also results in shear-induced dilation due to contacting asperities on the fault surface. Thus, fault permeability can increase by orders of magnitude (Lee and Cho, 2002; Guglielmi et al., 2015), leading to significant improvements in reservoir injectivity (Genter et al., 2010).

The extensively used Coulomb friction law incorporates the main mechanisms for reactivation and slip of faults. The Coulomb law gives an approximation of the threshold value for the frictional force on the surface (i.e., the contact shear traction) above which sliding occurs. Let $e_n$ denote the unit normal to the fracture surface in direction of the normal force exerted from the fracture to the matrix, and let $e_\tau$ denote the unit normal in the direction of the shear force on the matrix from the fracture. For a contact traction on the surface $t_c = t_c^n e_n + t_c^\tau e_\tau$, a Coulomb friction law is given by

$$t_c^\tau \leq \mu t_c^n + c, \quad (1)$$

where $\mu$ is the friction coefficient, and $c$ represents cohesion. The right-hand side of the equation represents the critical shear traction for sliding. Under crustal conditions, $c$ is normally negligible. Typical values for $\mu$ are between 0.3 and 1. The lower values apply for faults with clay fillings, which also have non-negligible cohesion. In general, $\mu$ is not constant, and calibration to shear experiments and observations of fault reactivation have introduced various empirical models for the friction coefficient depending on the time of contact, displacement, and sliding velocity (Dieterich, 1979; Ruina, 1983; Marone, 1998; Guglielmi et al., 2015). The richer models, however, face a risk of overparameterization considering limited available data and the associated uncertainty. Inherently, the models are inhibited by insufficient understanding of the physics of fracture deformation, and empirical relations may incorporate effects that could also have been modeled differently. An example is slip-strengthening friction models, which introduce strengthening of the fault with shear-induced dilatancy (Segall and Rice, 1995). The same effect will also be apparent in physics-based models correctly including dilation of faults with slip and corresponding stress redistribution in the surrounding formation, as dilatancy will result in increased normal load on the fault. In a recent study, Scuderi et al. (2017) conclude "that even for small changes in fluid pressure the effect of effective normal stress on fault strength and stability outweighs the rate and state dependent effects promoting fault unstable behavior," thereby supporting the need for models to correctly capture fluid pressure and poroelastic responses of slip and dilation.

The reactivation and displacement of faults will depend on how preexisting tractions on the faults are affected by hydromechanical stress changes on the fault, including alterations in fault pressure. To see this coupling, it is instructive to consider the force balance on the fault surface. Due to the roughness of the fracture surfaces, asperities on the fracture surfaces can keep the fracture hydraulically open while the two sides of the fracture are mechanically in contact. Hence, a combination of forces acts on the matrix: a contact force $\mathbf{t}_c$ resulting from contacting fracture asperities across the fault's surface and a normal force $p_f \mathbf{e}_n$ on the fault's surface exerted by the fluid pressure in the fracture (Hubbert and Rubey, 1959). This total force is balanced by the traction on the fracture surface from the matrix, $\mathbf{t}_m = t_m^n(-\mathbf{e}_n) + t_m^\tau(-\mathbf{e}_\tau)$, caused by hydromechanical stress. The force balance can be written as $-\mathbf{t}_m = \mathbf{t}_c + p_f \mathbf{e}_n$. Evaluating normal and shear forces, the friction law in equation 1 becomes

$$t_m^\tau \leq \mu\left(t_m^n - p_f\right) + c \ . \qquad (2)$$

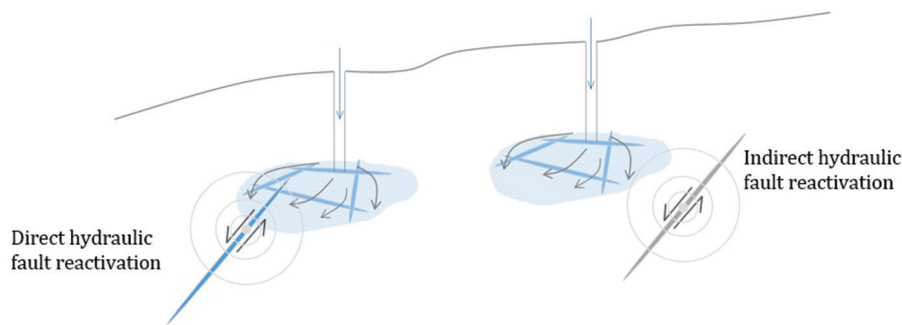

**Figure 1.** Illustration of fault reactivation in stimulation of a reservoir. The fault can be hydraulically reactivated directly through increased fluid pressure in the fault and indirectly through induced changes in the local thermo-poroelastic stress regime around the fault.

Based on this formula, a fracture is stable if the strict inequality is satisfied, while the fracture slips when the two sides of the equation become equal. A classical Mohr diagram (e.g., Jaeger et al., 2007) shows how the stability of a fault is related to its orientation and how both increase of fluid pressure in the fault and changes in hydromechanical stress on the fault can reactivate it (Figure 2). However, fluid injection does not always reduce fault strength; for example, if the fault is a barrier to fluid flow, increase in fluid pressure at one side of the fault will lead to an increase in the normal forces acting on the fault from the matrix and, hence, stabilize the fault.

Once a fault is reactivated, the slip is governed by equality in equation 2. By balance of forces in equation 2, coupled with the corresponding hydromechanical model for the surrounding domain discussed in the following section, displacement can be computed on the fault. In hard rocks, fault slip is associated with dilation as sliding alters the contact of asperities and moves the fracture's two sides farther apart. The increase in hydraulic aperture is typically proportional to slip up to a threshold value where production of gauge can explain the prevention of further permeability increase. Empirical relationships between slip and dilation, as well as the effect of dilatancy on fault permeability, are studied by, e.g., Lee and Cho (2002). Furthermore, this fault dilation interacts with the hydromechanical stresses on the fault, leading to a significant coupling (Stefansson et al., 2020).

## Modeling of fault reactivation and slip accounting for process-structure interaction

It is crucial for physics-based simulation models to capture the process-structure interaction discussed in the previous section. This requires models that (1) explicitly represent dominating fault structures and (2) couple flow, reactivation, and deformation of the faults with flow and hydromechanical deformation of the formation where the faults reside (matrix).

The matrix surrounding the explicitly represented faults needs, in general, to be modeled as porous and permeable due to its incorporation of finer-scale fractures and porous rock (Berre et al., 2019). Significant progress has been made toward the development of models based on the conceptual principles 1 and 2 listed in the preceding paragraph. While early models did not consider redistribution of stress as a consequence of slip (Willis-Richards et al., 1996; Rahman et al., 2002; Bruel, 2007; Kohl and Mégel, 2007), simplified models for stress redistribution based on block-spring models (Baisch et al., 2010), the semianalytical boundary integral method (Ghassemi and Zhou, 2011; McClure and Horne, 2011; Norbeck et al., 2016), and/or simplified models for approximating fracture slip in the coupled problem (Ucar et al., 2018) were later introduced. However, neither of these models consider the hydromechanical (poroelastic) deformation of the matrix surrounding the faults, and only Ucar et al. (2018) model the full 3D momentum balance for elastic deformation of the matrix coupled with normal and shear deformation of faults.

Based on conservation principles and constitutive relations, the conceptual principles 1 and 2 result in a coupled system of partial differential equations governing fault reactivation and deformation. Recently, numerical models based on discretization and fully coupled solutions of the governing system of equations have been developed (Garipov et al., 2018; Garipov and Hui, 2019; Berge et al., 2020; Keilegavlen et al., 2020; Stefansson et al., 2020), which also show appropriate grid convergence properties (Stefansson et al., 2020). These models are based on conceptualizing the faults as 2D surfaces in the 3D domain. This avoids elements with large aspect ratios on the fault (Karimi-Fard et al., 2003) and facilitates modeling of slip and dilation (Cappa and Rutqvist, 2011; Ucar et al., 2018). We have based the model presented in the following on this principle.

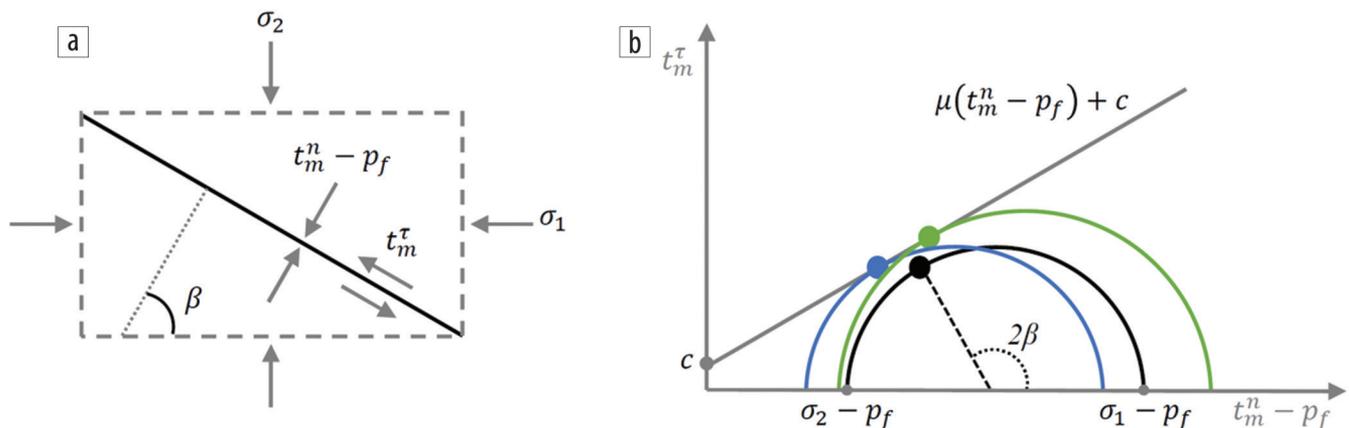

**Figure 2**. A fracture oriented at an angle $\beta$ to the maximum principal stress direction is affected by normal and shear contact forces given by the background hydromechanical stress field and fracture fluid pressure. (a) Illustration of fracture and the normal and shear contact forces acting on it. (b) Mohr diagram illustrating the Coulomb friction law. From an initial state of the forces acting on the fracture surface (black dot), the fracture can be reactivated by an increase of fluid pressure (blue dot) or a change in the hydromechanical forces from the matrix on the fracture surface (green dot). The stress states of fractures with other orientations subjected to the same principal stresses and fluid pressures as the mentioned states are illustrated by the half circles of corresponding colors.

In the mathematical model, the slip along each fault is governed by the constitutive Coulomb friction law (equation 2). Furthermore, the governing system of equations incorporates fluid flow in the faults and hydromechanics in the matrix. For this, flow in both faults and matrix is assumed to be single-phase and governed by Darcy's law, and the matrix deformation is assumed to be poroelastic. Shear-induced dilation is governed by a constitutive law that relates mechanical aperture increase to slip distance along a fault through a dilation angle. Based on the simplifying assumption that the hydraulic and mechanical aperture of a permeable fault are the same, transmissivity of a permeable fault is coupled to its aperture through the cubic law. Excluding thermal effects, we use the same full set of model equations as Stefansson et al. (2020).

## Mechanisms affecting fault reactivation and slip through coupled processes

In this section, we utilize the advantages of physics-based modeling and discuss aspects affecting fault reactivation and slip illustrated by numerical examples. The test cases are set up using the open-source software PorePy (Keilegavlen et al., 2020) including the recent developments by Stefansson et al. (2020). The open-source software as well as all details of the simulations that are the basis for the presented results, including run scripts for the examples, are available at GitHub (Stefansson, 2020).

We have designed the examples to illustrate several of the mechanisms discussed earlier based on a rather simple geometry. The domain is located with its top at the earth's surface and extends 2 km in the vertical direction and in both horizontal directions. Two square faults (faults 1 and 2) are located in the center of the domain as shown in Figure 3, both with an extension of 170 m in both directions along the fault plane. The boundary conditions on the stress are given by a background stress with principal axis aligned with the coordinate system. Vertical stress, $\sigma_V = \sigma_z$, is set as lithostatic, and principal horizontals stresses are $\sigma_H = \sigma_x = 1.3\sigma_z$ and $\sigma_h = \sigma_y = 0.6\sigma_z$, respectively. For pressure, boundary conditions are given as hydrostatic.

Table 1 lists the parameters for the simulation of a reference case (case 0). For simplicity, the fault aperture, $a$, equals both mechanical and hydraulic aperture, and, in the case of permeable faults, an isotropic fault permeability is given by the cubic law; that is, it equals $a^2/12$. The domain is meshed with 27,215 cells in the matrix and, in total, 1092 cells in the faults.

Assuming zero cohesion, $c$, a measure of each fault's tendency to slip based on equation 2 is given by $t_m^\tau/(t_m^n - p_f)$. Before injection starts, the slip tendency of fault 1 is 0.18, implying that the fault is far from being critically stressed; fault 2 has a slip tendency of 0.47, which is close to the frictional threshold of 0.5 for slip.

From the described initial state, water is injected into fault 1 (Figure 3) at a rate of 60 l/s. The hydraulic stimulation is done for five days. To illustrate different mechanisms in fault reactivation and slip due to the fluid injection, we introduce three cases, each showing the result of changing one central parameter compared to case 0. In case A, the matrix permeability is changed from the case 0 value of $2.5 \cdot 10^{-15}$ m$^2$ to $4.0 \cdot 10^{-15}$ m$^2$. In case B, the permeability of fault 2 is set to the constant value of $1.0 \cdot 10^{-18}$ m$^2$ during the entire simulation, resulting in fault 2 effectively being a barrier to flow. In case C, the Biot coefficient is reduced from 0.7 to 0.6. For each case, we show the following results at the end of the stimulation: slip tendency (Figure 4), fluid pressure (Figure 5), and total aperture increase (Figure 6), which is proportional to slip distance along the fault.

For case 0, both fault 1 and fault 2 are destabilized as a consequence of the fluid injection. In fault 1, the increase in pressure (Figure 4) increases slip tendency along the fault (Figure 5) and even results in slip along some elements at the top of the fault (Figure 6). For fault 2, which was almost critically stressed initially, the entire fault surface has slipped by the end of the stimulation. In this case, only relatively small changes in pressure and poroelastic stress conditions were necessary to induce slip.

**Table 1.** Rock and fluid parameters for base case.

| Bulk modulus | $2.2 \cdot 10^{10}$ Pa |
|---|---|
| Shear modulus | $1.7 \cdot 10^{10}$ Pa |
| Solid density | $2.7 \cdot 10^{3}$ kg/m$^3$ |
| Biot coefficient | 0.7 |
| Porosity | $1.0 \cdot 10^{-2}$ |
| Matrix permeability | $2.5 \cdot 10^{-15}$ m$^2$ |
| Viscosity | $1.0 \cdot 10^{-3}$ Pa s |
| Fluid compressibility | $1.0 \cdot 10^{-10}$ 1/Pa |
| Fluid reference density | $1.0 \cdot 10^{3}$ kg/m$^3$ |
| Initial fault aperture | $1.0 \cdot 10^{-3}$ m |
| Friction coefficient ($\mu$) | 0.5 |
| Cohesion ($c$) | 0 |
| Dilation angle | 5° |

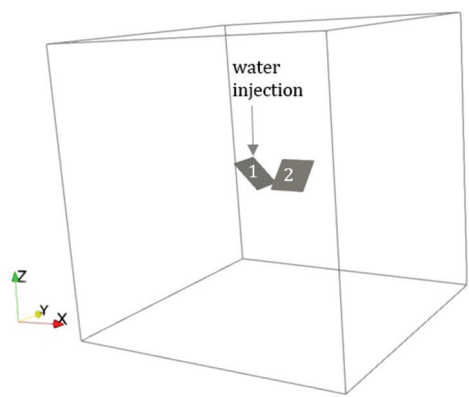

**Figure 3.** Fault geometry of faults 1 and 2 (numbered in the figure) and computational domain. Water is injected in one of the top corners of fault 1, illustrated with an arrow.

The higher matrix permeability in case A compared to case 0 allows the pressure front to diffuse more easily into the matrix. This gives a reduced pressure in fault 1 compared to case 0 (Figure 4) and, consequently, a weaker increase in slip tendency than was seen in case 0. At the same time, the pressure in fault 2 is increased sufficiently to cause slip of the entire fault but with lower slip magnitudes and corresponding aperture changes than in case 0.

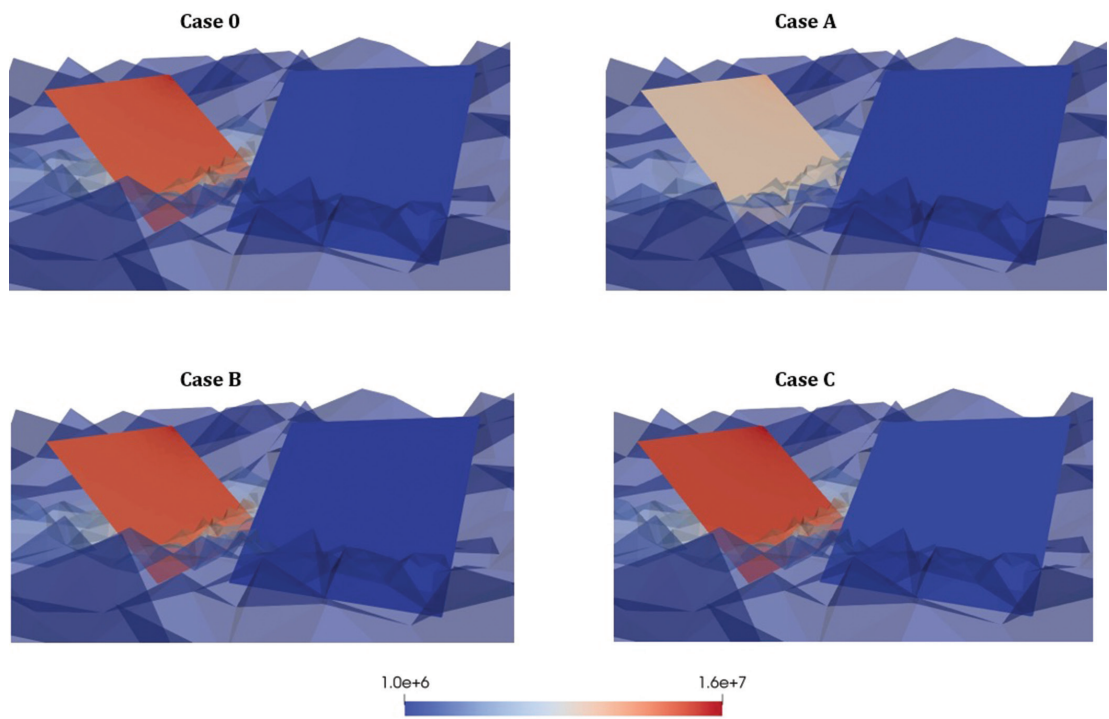

**Figure 4.** Fluid pressure difference (Pa) to initial hydrostatic pressure for case 0, case A (higher matrix permeability), case B (fault 2 blocking), and case C (lower Biot coefficient). The matrix pressure is shown for a cross-section of the domain in a transparent manner, also indicating the refinement of the grid toward the faults.

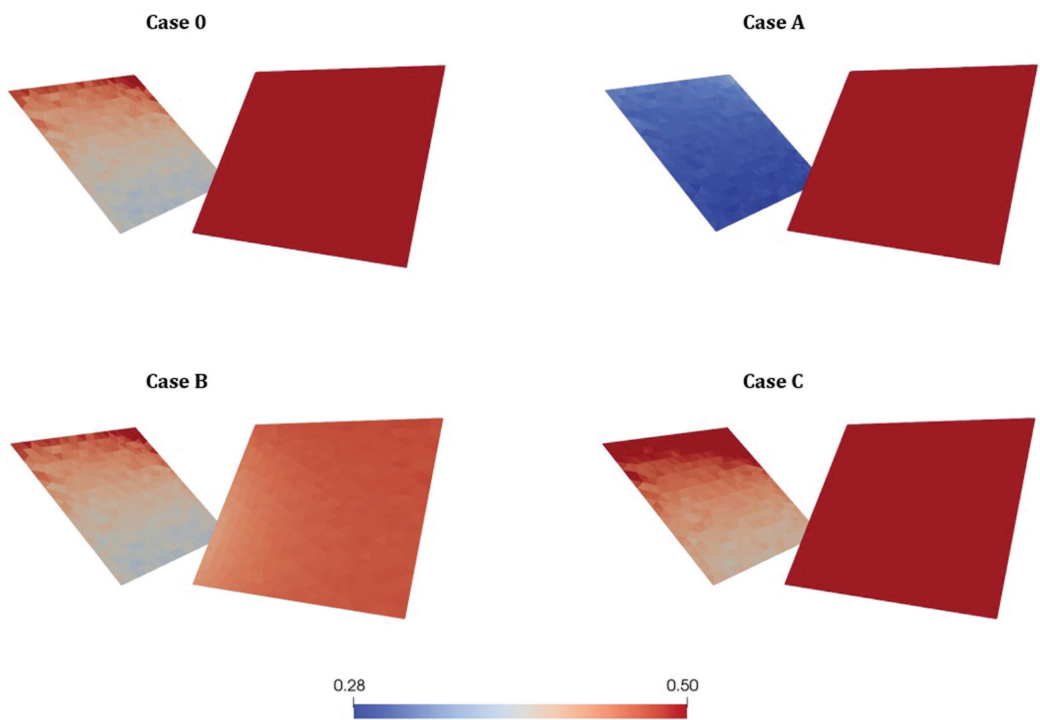

**Figure 5.** Slip tendency of faults for case 0, case A (higher matrix permeability), case B (fault 2 blocking), and case C (lower Biot coefficient). Elements that are slipping have the maximum slip tendency of 0.5.

Special Section: Geothermal energy                                December 2020    **The Leading Edge**    897

Compared to the initial situation before stimulation, case B results in a stabilization of fault 2 as the fluid pressure results in increased loading on the fault due to it effectively being a barrier for the flow. This is recognized by a reduction in slip tendency of the fault from the initial slip tendency of 0.47 (Figure 5).

The lower Biot coefficient of case C compared to case 0 results in a weaker Biot coupling. The formation is stiffer, and elastically contributes less to the migration of pressure out of fault 1. In case C, fault 1 has a pressure difference to hydrostatic pressure ranging between 14.3 and 15.7 MPa as compared to 13.8 and 15.3 MPa for case 0. The result is an increased slip tendency along fault 1 compared to case 0. Compared to case 0, significantly larger parts of the fault slip. Fault 2 also has more slip and corresponding aperture increase than it has in case 0. This happens despite the pressure being only slightly increased in fault 2 compared to case 0. In case C, fault 2 has a pressure difference to hydrostatic pressure ranging between 1.42 and 1.45 MPa larger than hydrostatic pressure as compared to 1.22 and 1.25 MPa for case 0. This indicates that poroelastic stress changes resulting from the slip of fault 1 have a significant impact on the slip along fault 2. However, even for this simple geometry and without parameter heterogeneities in the matrix, the processes are so strongly coupled that it is difficult to distinguish which mechanisms cause slip along the different faults.

## Conclusions

The hydromechanical processes in hydraulic stimulation of a geothermal reservoir by fault reactivation and slip are coupled and interact with the deforming fractured structure of the formation. Numerical modeling tools must acknowledge the full coupling between processes as well as fracture reactivation, slip, and dilation, which affect the fracture's aperture and consequently its influence on flow. A methodology has been developed recently that fully couples flow in faults and matrix with poroelastic matrix deformation and a contact mechanics model for the faults, including fault dilation because of slip. In this paper, we have shown how an open-source simulation tool that implements this methodology can be used to assess the effect of various subsurface characteristics and mechanisms for fracture reactivation and slip.

The numerical results demonstrate how even relatively small changes in rock and fault parameters can affect fault deformation significantly. Furthermore, they show that the coupling in the hydromechanical process-structure interaction is so strong that in several situations it becomes impossible to identify simple causality. Hence, models such as the one presented herein, accounting for the fully coupled physics, enable assessments of dynamics that would have been impossible otherwise. Combined with field and experimental data, they have potential to improve our understanding of governing mechanisms in hydraulic stimulation of geothermal reservoirs and to forecast outcomes of engineering operations based on various subsurface scenarios. **TLE**

## Acknowledgments

The work was funded by the Research Council of Norway through grant no. 267908 and Equinor ASA.

## Data and materials availability

Data associated with this research are available and can be accessed via the following URL: https://github.com/IvarStefansson/Fault-Slip-in-Hydraulic-Stimulation-of-Geothermal-Reservoirs.

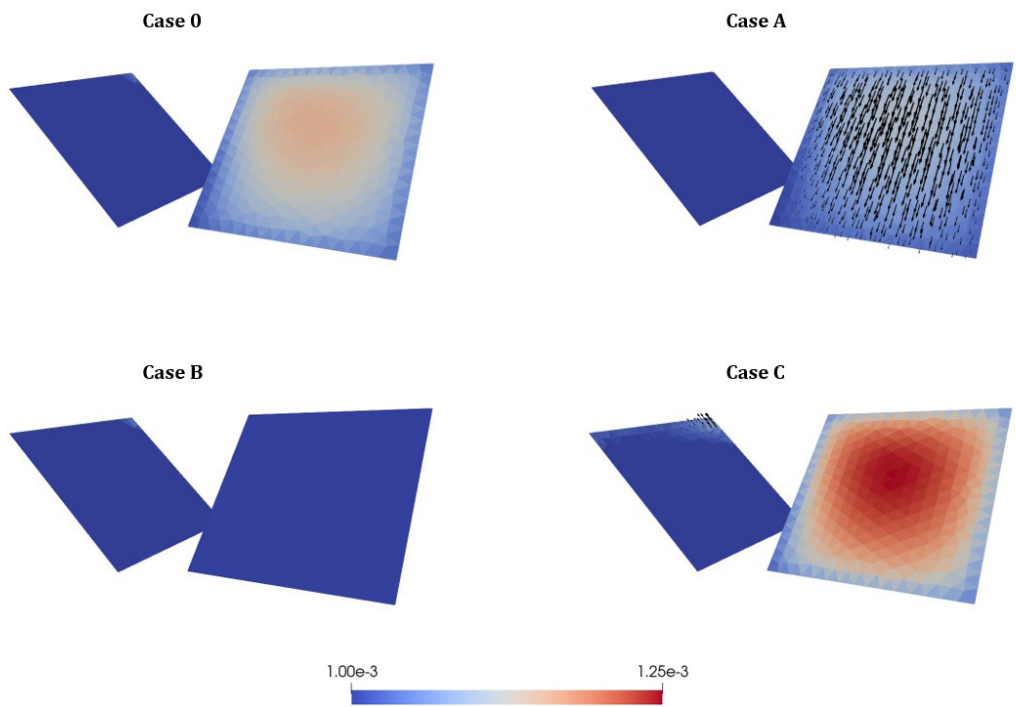

**Figure 6.** Aperture of faults (mm) after hydraulic stimulation for case 0, case A (higher matrix permeability), case B (fault 2 blocking), and case C (lower Biot coefficient). Arrows illustrating the magnitude and direction of slip for the top surfaces of the faults are included for fault 2, case A and fault 1, case C to illustrate the direction and magnitude of slip.

Corresponding author: inga.berre@uib.no

## References

Baisch, S., R. Vörös, E. Rothert, H. Stang, R. Jung, and R. Schellschmidt, 2010, A numerical model for fluid injection induced seismicity at Soultz-sous-Forêts: International Journal of Rock Mechanics and Mining Sciences, **47**, no. 3, 405–413, https://doi.org/10.1016/j.ijrmms.2009.10.001.

Berge, R. L., I. Berre, E. Keilegavlen, J. M. Nordbotten, and B. Wohlmuth, 2020, Finite volume discretization for poroelastic media with fractures modeled by contact mechanics: International Journal for Numerical Methods in Engineering, **121**, no. 4, 644–663, https://doi.org/10.1002/nme.6238.

Berre, I., F. Doster, and E. Keilegavlen, 2019, Flow in fractured porous media: A review of conceptual models and discretization approaches: Transport in Porous Media, **130**, no. 1, 215–236, https://doi.org/10.1007/s11242-018-1171-6.

Bethmann, F., N. Deichmann, and P. M. Mai, 2011, Scaling relations of local magnitude versus moment magnitude for sequences of similar earthquakes in Switzerland: Bulletin of the Seismological Society of America, **101**, no. 2, 515–534, https://doi.org/10.1785/0120100179.

Breede, K., K. Dzebisashvili, X. Liu, and G. Falcone, 2013, A systematic review of enhanced (or engineered) geothermal systems: Past, present and future: Geothermal Energy, **1**, article no. 4, https://doi.org/10.1186/2195-9706-1-4.

Bruel, D., 2007, Using the migration of the induced seismicity as a constraint for fractured hot dry rock reservoir modelling: International Journal of Rock Mechanics and Mining Sciences, **44**, no. 8, 1106–1117, https://doi.org/10.1016/j.ijrmms.2007.07.001.

Cappa, F., and J. Rutqvist, 2011, Modeling of coupled deformation and permeability evolution during fault reactivation induced by deep underground injection of $CO_2$: International Journal of Greenhouse Gas Control, **5**, no. 2, 336–346, https://doi.org/10.1016/j.ijggc.2010.08.005.

Deichmann, N., and D. Giardini, 2009, Earthquakes induced by the stimulation of an enhanced geothermal system below Basel (Switzerland): Seismological Research Letters, **80**, no. 5, 784–798, https://doi.org/10.1785/gssrl.80.5.784.

Dieterich, J. H., 1979, Modeling of rock friction: 1. Experimental results and constituve equations: Journal of Geophysical Research, **84**, no. 9, 2161–2168, https://doi.org/10.1029/JB084iB05p02161.

Eisner, L., P. M. Duncan, W. M. Heigl, and W. R. Keller, 2009, Uncertainties in passive seismic monitoring: The Leading Edge, **28**, no. 6, 648–655, https://doi.org/10.1190/1.3148403.

Ellsworth, W. L., 2013, Injection-induced earthquakes: Science, **341**, no. 6142, https://doi.org/10.1126/science.1225942.

Garipov, T. T., and M. H. Hui, 2019, Discrete fracture modeling approach for simulating coupled thermo-hydro-mechanical effects in fractured reservoirs: International Journal of Rock Mechanics and Mining Sciences, **122**, https://doi.org/10.1016/J.IJRMMS.2019.104075.

Garipov, T. T., P. Tomin, R. Rin, D. V. Voskov, and H. A. Tchelepi, 2018, Unified thermo-compositional-mechanical framework for reservoir simulation: Computational Geosciences, **22**, no. 4, 1039–1057, https://doi.org/10.1007/s10596-018-9737-5.

Genter, A., K. Evans, N. Cuenot, D. Fritsch, and B. Sanjuan, 2010, Contribution of the exploration of deep crystalline fractured reservoir of Soultz to the knowledge of enhanced geothermal systems (EGS): Comptes Rendus Geoscience, **342**, no. 7–8, 502–516, https://doi.org/10.1016/J.CRTE.2010.01.006.

Geological Society of Korea, 2019, Summary report of the Korean Government Commission on relations between the 2017 Pohang earthquake EGS project: https://doi.org/10.22719/KETEP-20183010111860.

Ghassemi, A., 2012, A review of some rock mechanics issues in geothermal reservoir development: Geotechnical and Geological Engineering, **30**, no. 3, 647–664, https://doi.org/10.1007/s10706-012-9508-3.

Ghassemi, A., and X. Zhou, 2011, A three-dimensional thermo-poroelastic model for fracture response to injection/extraction in enhanced geothermal systems: Geothermics, **40**, no. 1, 39–49, https://doi.org/10.1016/J.GEOTHERMICS.2010.12.001.

Goebel, T. H. W., M. Weingarten, X. Chen, J. Haffener, and E. E. Brodsky, 2017, The 2016 Mw5.1 Fairview, Oklahoma earthquakes: Evidence for long-range poroelastic triggering at >40 km from fluid disposal wells: Earth and Planetary Science Letters, **472**, 50–61, https://doi.org/10.1016/j.epsl.2017.05.011.

Guglielmi, Y., F. Cappa, J. P. Avouac, P. Henry, and D. Elsworth, 2015, Seismicity triggered by fluid injection-induced aseismic slip: Science, **348**, no. 6240, 1224–1226, https://doi.org/10.1126/science.aab0476.

Herrmann, R. B., S.-K. Park, and C.-Y. Wang, 1981, The Denver earthquakes of 1967–1968: Bulletin of the Seismological Society of America, **71**, no. 3, 731–745.

Hsieh, P. A., and J. D. Bredehoeft, 1981, A reservoir analysis of the Denver earthquakes: A case of induced seismicity: Journal of Geophysical Research: Solid Earth, **86**, no. B2, 903–920, https://doi.org/10.1029/JB086iB02p00903.

Hubbert, M. K., and W. W. Rubey, 1959, Role of fluid pressure in mechanics of overthrust faulting: I. Mechanics of fluid-filled porous solids and its application to overthrust faulting: GSA Bulletin, **70**, no. 2, 115–166, https://doi.org/10.1130/0016-7606(1959)70[115:rofpim]2.0.co;2.

Jaeger, J. C., N. G. W. Cook, and R. W. Zimmerman, 2007, Fundamentals of rock mechanics, 4$^{th}$ edition: Blackwell Publishing.

Karimi-Fard, M., L. J. Durlofsky, and K. Aziz, 2003, An efficient discrete fracture model applicable for general purpose reservoir simulators: SPE Reservoir Simulation Symposium, Society of Petroleum Engineers, https://doi.org/10.2118/79699-MS.

Keilegavlen, E., R. Berge, A. Fumagalli, M. Starnoni, I. Stefansson, J. Varela, and I. Berre, 2020, PorePy: An open-source software for simulation of multiphysics processes in fractured porous media: Computational Geoscience, https://doi.org/10.1007/s10596-020-10002-5.

Kohl, T., and T. Mégel, 2007, Predictive modeling of reservoir response to hydraulic stimulations at the European EGS site Soultz-sous-Forêts: International Journal of Rock Mechanics and Mining Sciences, **44**, no. 8, 1118–1131, https://doi.org/10.1016/j.ijrmms.2007.07.022.

Lee, H. S., and T. F. Cho, 2002, Hydraulic characteristics of rough fractures in linear flow under normal and shear load: Rock Mechanics and Rock Engineering, **35**, no. 4, 299–318, https://doi.org/10.1007/s00603-002-0028-y.

Lee, K. K., W. Ellsworth, D. Giardini, J. Townend, S. Ge, T. Shimamoto, I.-W. Yeo, et al., 2019, Managing injection-induced seismic risks: Science, **364**, no. 6442, 730–732, https://doi.org/10.1126/science.aax1878.

Marone, C., 1998, Laboratory-derived friction laws and their application to seismic faulting: Annual Review of Earth and Planetary Sciences, **26**, 643–696, https://doi.org/10.1146/annurev.earth.26.1.643.

McClure, M. W., and R. N. Horne, 2011, Investigation of injection-induced seismicity using a coupled fluid flow and rate/state friction model: Geophysics, **76**, no. 6, WC181–WC198, https://doi.org/10.1190/geo2011-0064.1.

Moeck, I. S., 2014, Catalog of geothermal play types based on geologic controls: Renewable and Sustainable Energy Reviews, **37**, 867–882, https://doi.org/10.1016/J.RSER.2014.05.032.

Norbeck, J. H., M. W. McClure, J. W. Lo, and R. N. Horne, 2016, An embedded fracture modeling framework for simulation of hydraulic fracturing and shear stimulation: Computational Geosciences, **20**, no. 1, 1–18, https://doi.org/10.1007/s10596-015-9543-2.

Pine, R. J., and A. S. Batchelor, 1984, Downward migration of shearing in jointed rock during hydraulic injections: International Journal of Rock Mechanics and Mining Sciences & Geomechanics Abstracts, **21**, no. 5, 249–263, https://doi.org/10.1016/0148-9062(84)92681-0.

Rahman, M. K., M. M. Hossain, and S. S. Rahman, 2002, A shear-dilation-based model for evaluation of hydraulically stimulated naturally fractured reservoirs: International Journal for Numerical and Analytical Methods in Geomechanics, **26**, no. 5, 469–497, https://doi.org/10.1002/nag.208.

Ruina, A., 1983, Slip instability and state variable friction laws: Journal of Geophysical Research: Solid Earth, **88**, no. B12, 10359–10370, https://doi.org/10.1029/JB088iB12p10359.

Scuderi, M. M., C. Collettini, and C. Marone, 2017, Frictional stability and earthquake triggering during fluid pressure stimulation of an experimental fault: Earth and Planetary Science Letters, **477**, 84–96, https://doi.org/10.1016/j.epsl.2017.08.009.

Segall, P., and J. R. Rice, 1995, Dilatancy, compaction, and slip instability of a fluid-infiltrated fault: Journal of Geophysical Research: Solid Earth, **100**, no. B11, 22155–22171, https://doi.org/10.1029/95JB02403.

Stefansson, I., 2020, Run scripts for the paper "Fault slip in hydraulic stimulation of geothermal reservoirs: Governing mechanisms and process-structure interaction" by I. Berre, E. Keilegavlen, and I. Stefansson: https://github.com/IvarStefansson/Fault-Slip-in-Hydraulic-Stimulation-of-Geothermal-Reservoirs.

Stefansson, I., I. Berre, and E. Keilegavlen, 2020, A fully coupled numerical model of thermo-hydro-mechanical processes and fracture contact mechanics in porous media: arxiv.org/abs/2008.06289

Ucar, E., I. Berre, and E. Keilegavlen, 2018, Three-dimensional numerical modeling of shear stimulation of fractured reservoirs: Journal of Geophysical Research: Solid Earth, **123**, no. 5, 3891–3908, https://doi.org/10.1029/2017JB015241.

Willis-Richards, J., K. Watanabe, and H. Takahashi, 1996, Progress toward a stochastic rock mechanics model of engineered geothermal systems: Journal of Geophysical Research: Solid Earth, **101**, no. B8, 17481–17496, https://doi.org/10.1029/96JB00882.

Wintsch, R. P., R. Christoffersen, and A. K. Kronenberg, 1995, Fluid-rock reaction weakening of fault zones: Journal of Geophysical Research: Solid Earth, **100**, no. B7, 13021–13032, https://doi.org/10.1029/94JB02622.

Zang, A., V. Oye, P. Jousset, N. Deichmann, R. Gritto, A. McGarr, E. Majer, and D. Bruhn, 2014, Analysis of induced seismicity in geothermal reservoirs — An overview: Geothermics, **52**, 6–21, https://doi.org/10.1016/J.GEOTHERMICS.2014.06.005.